\documentclass[10pt, conference, letterpaper]{IEEEtran}
\usepackage[letterpaper, left=0.65in, right=0.65in, bottom=1.01in, top=0.75in]{geometry}
\IEEEoverridecommandlockouts
\usepackage{cite}
\usepackage{amsmath,amssymb,amsfonts}
\usepackage{graphicx}
\usepackage[caption=false,font=footnotesize]{subfig}
\usepackage{textcomp}
\usepackage{xcolor}
\usepackage{tabularx}
\usepackage{multirow}
\usepackage{diagbox}
\usepackage[binary-units]{siunitx} 
\DeclareSIUnit{\bps}{bps}
\usepackage{amsthm}
\newtheorem{proposition}{Proposition}
\newtheorem{remark}{Remark}
\usepackage[ruled,vlined]{algorithm2e}
\usepackage{arydshln}

\begin{document}

\title{What is the Most Efficient Technique for Uplink Cell-Free Massive MIMO?}

\author{\IEEEauthorblockN{Wei Jiang\IEEEauthorrefmark{1} and Hans D. Schotten\IEEEauthorrefmark{2}}
\IEEEauthorblockA{\IEEEauthorrefmark{1}German Research Center for Artificial Intelligence (DFKI)\\Trippstadter Street 122,  Kaiserslautern, 67663 Germany\\
  }
\IEEEauthorblockA{\IEEEauthorrefmark{2}Technische Universit\"at  (RPTU) Kaiserslautern\\Building 11, Paul-Ehrlich Street, Kaiserslautern, 67663 Germany\\
 }
\thanks{This work was supported by the German Federal Ministry of Education and Research (BMBF) through \emph{Open6G-Hub} (Grant no.  \emph{16KISK003K}) research project.}
}
\maketitle

\begin{abstract}
This paper seeks to determine the most efficient uplink technique for cell-free massive MIMO systems. Despite offering great advances, existing works suffer from fragmented methodologies and inconsistent assumptions (e.g., single- vs. multi-antenna access points, ideal vs. spatially correlated channels). To address these limitations, we: (1) establish a unified analytical framework compatible with centralized/distributed processing and diverse combining schemes; (2) develop a universal optimization strategy for max-min power control; and (3) conduct a holistic study among four critical metrics: worst-case user spectral efficiency (fairness), system capacity, fronthaul signaling, and computational complexity. Through analyses and evaluation, this work ultimately identifies the optimal uplink technique for practical cell-free deployments.  
\end{abstract}

\section{Introduction}

In traditional cellular networks, users near the base station generally enjoy high quality of service (QoS), while users at the cell edge suffer from low QoS due to weak signal strength and strong inter-cell interference \cite{Ref_jiang2024TextBook}. One of the most critical missions for next-generation networks is to really meet the demand for ubiquitous high-quality connectivity. Cell-free massive multiple-input multiple-output (CF-mMIMO) has emerged as a transformative paradigm in cellular networks, providing uniformly high spectral efficiency (SE)  by eliminating cellular boundaries. In CF-mMIMO, numerous access points (APs) are distributed over a wide area and jointly serve users on the same time-frequency resources, coordinated by a central processing unit (CPU).

In the uplink of CF-mMIMO, the critical aspects of transmission strategies—encompassing signal detection \cite{Zhang2024LSFD}, power control \cite{Ref_bashar2019uplink}, fronthaul signaling \cite{Bashar2020ExploitingDL}, and computational complexity~\cite{Ref_bjornson2019scalable} —play a pivotal role in determining system performance and deployment costs. Despite extensive research works, this field lacks a cohesive analysis and evaluation   due to fragmented methodologies and inconsistent assumptions across studies. For instance, some approaches rely on centralized methods that exploit global channel state information (CSI) at the CPU, while others adopt decentralized strategies, distributing processing tasks across APs. Additionally, many earlier investigations assumed APs with a single antenna and independent fading channels to simplify their models, though recent works have tackled the more practical scenario of multi-antenna APs with spatially correlated channels \cite{Ref_wang2021uplink, Ref_bjornson2020making}. Moreover, while much of the research emphasizes worst-case performance (fairness), other critical factors—such as total system capacity, fronthaul overhead, and computational complexity—also demand attention. To date, the absence of a unified analytical framework and a comprehensive trade-off comparison leaves a question: \textit{what is the most efficient uplink technique for cell-free massive MIMO?}

In this paper, we seek to provide the answer by implementing the following strategies:
\begin{itemize}
    \item We provide a unified SE analysis that accommodates any number of antennas per AP, integrates spatially correlated channel models, and considers pilot contamination. This analytical framework supports various combining methods, including maximal-ratio (MR), zero-forcing (ZF), regularized ZF (RZF), minimum mean-square error (MMSE), and large-scale fading decoding (LSFD), for both centralized and decentralized architectures.
    \item We propose a universal max-min power optimization algorithm applicable to all combining methods.
    \item We make a comprehensive study about trade-offs among four critical metrics: (i) worst-case user SE, (ii) total system capacity, (iii) fronthaul signaling, and (iv) computational complexity. 
\end{itemize}

The structure of the paper is as follows: Section II introduces the system model. Sections III and IV develop the analytical framework for centralized and decentralized processing, respectively. Sections V to VII focus on power optimization, computational complexity, and fronthaul overhead, respectively. Section VIII offers simulation results, and Section IX concludes the work.

\section{System Model}

In CF-mMIMO, a network of $L$ APs, each equipped with $N_a$ co-located antennas, are distributed across a coverage area to serve $K$ users. To exploit the advantages of channel hardening and favorable propagation, the total number of service antennas $M=L\times N_a$ must substantially exceed the number of users, i.e., $M\gg K$. User equipment (UE) is typically assumed to be equipped with a single antenna. The index sets of APs and users are denoted by $\mathbb{L}= \{1,\ldots,L\}$ and $\mathbb{K}=\{1,\ldots,K\}$, respectively.

The channel between AP $l$, $\forall l\in \mathbb{L}$ and user $k$, $\forall k \in \mathbb{K}$ is denoted by $\mathbf{h}_{kl}\in \mathbb{C}^{N_a}$. 
Under the standard block fading model, each \textit{coherent block} is a time-frequency interval of $\tau_c$ channel uses during which the channel response remains approximately constant. Most prior works on CF-mMIMO have simplified analyses by assuming either single-antenna APs or multi-antenna APs with independent channels. However, in practical scenarios, closely spaced antennas at a multi-antenna AP exhibit correlated channel responses. Thus,  each coherence block applies an independent realization from \textit{correlated} Rayleigh fading, defined as $\mathbf{h}_{kl} \sim \mathcal{CN}(\mathbf{0}, \mathbf{R}_{kl} )$, where $\mathbf{R}_{kl}=\mathbb{E}[  \mathbf{h}_{kl} \mathbf{h}_{kl}^H ]$ 
stands for the spatial correlation matrix \cite{yu2004modeling}.

Through linear MMSE channel estimation \cite{Ref_bjornson2020making}, the estimate of $\mathbf{h}_{kl}$ follows a complex Gaussian distribution:
\begin{equation}
    \hat{\mathbf{h}}_{kl} \sim \mathcal{CN}(\mathbf{0}, p_u \tau_p \mathbf{R}_{kl} \boldsymbol{\Gamma}_{kl}^{-1}\mathbf{R}_{kl} ),
\end{equation}
where \begin{equation}
     \boldsymbol{\Gamma}_{kl}=p_u\tau_p\sum \nolimits_{k'\in \mathcal{P}_k } \mathbf{R}_{k'l} + \sigma_z^2\mathbf{I}_{N_a},
     \end{equation} 
and $p_u$ is UE's transmit power constraint, $\tau_p$ is the length of pilot sequences, $\mathcal{P}_k$ denotes the set of indices for the users, which utilize the same pilot sequence as user $k$, and $\sigma_z^2$ is the noise power.
The estimation error, defined as $\tilde{\mathbf{h}}_{kl}=\mathbf{h}_{kl} - \hat{\mathbf{h}}_{kl}$, arises due to additive noise and pilot contamination, following $\mathcal{CN}(\mathbf{0}, \boldsymbol{\Theta }_{kl} )$, where its correlation matrix is given by
\begin{equation}
\boldsymbol{\Theta}_{kl}=\mathbb{E}\left[ \tilde{ \mathbf{h}}_{kl} \tilde{\mathbf{h}}_{kl}^H \right]=\mathbf{R}_{kl} - p_u \tau_p \mathbf{R}_{kl} \boldsymbol{\Gamma}_{kl}^{-1}\mathbf{R}_{kl}. 
\end{equation}

\section{Centralized Uplink Combining}
During uplink transmission, all UEs concurrently send their data symbols. Specifically, UE \( k \) transmits symbol \( x_k \) scaled by a power coefficient \( 0 \leq \eta_k \leq 1 \). These symbols exhibit zero mean, unit variance, and mutual statistical independence, meaning the covariance matrix of the symbol vector \( \mathbf{x} = [x_1, \ldots, x_K]^T \) satisfies \( \mathbb{E}[\mathbf{x}\mathbf{x}^H] = \mathbf{I}_K \). At each channel use, AP \( l \) receives the signal:
\begin{equation} \label{GS_uplink_RxsignalAP}
    \mathbf{y}_l = \sqrt{p_u} \sum\nolimits_{k \in \mathbb{K}} \sqrt{\eta_k} \mathbf{h}_{kl} x_k + \mathbf{z}_l,
\end{equation}
where \( \mathbf{z}_l \sim \mathcal{CN}(\mathbf{0}, \sigma_z^2 \mathbf{I}_{N_a}) \) represents the additive noise at the AP. For centralized processing, every AP \( l \) sends both \( \mathbf{y}_l \) and its collected pilot signals to the CPU. The CPU subsequently performs joint channel estimation and signal detection.

\subsection{Unified Analytical Framework for SE}
The received signal available at the CPU is collectively expressed by $\mathbf{y} = [\mathbf{y}_0^T,\mathbf{y}_1^T\ldots,\mathbf{y}_L^T]^T$, and we have
\begin{equation} \label{eqnUPLINKmodel} \nonumber
    \mathbf{y} = \sqrt{p_u} \sum_{k\in \mathbb{K}} \sqrt{\eta_k} \mathbf{h}_k x_k + \mathbf{z},
\end{equation}
where the receiver noise $\mathbf{z} = [\mathbf{z}_0^T,\mathbf{z}_1^T\ldots,\mathbf{z}_L^T]^T$. The channel signature for user $k$, that is, the overall channel coefficients between all service antennas to this user, is denoted by $\mathbf{h}_k = [\mathbf{h}_{k0}^T,\mathbf{h}_{k1}^T\ldots,\mathbf{h}_{kL}^T]^T$.
It is distributed as $\mathbf{h}_k \sim \mathcal{CN}(\mathbf{0}, \mathbf{R}_k )$, where $\mathbf{R}_k=\mathrm{diag} (\mathbf{R}_{k0},\mathbf{R}_{k1},\ldots,\mathbf{R}_{kL})$ is the block diagonal spatial correlation matrix. Correspondingly, the correlation matrix of estimation error $\tilde{\mathbf{h}}_{k}$ equals
\begin{align} \nonumber
    \boldsymbol{\Theta}_k & =\mathrm{diag} (\boldsymbol{\Theta}_{k0},\boldsymbol{\Theta}_{k1},\ldots,\boldsymbol{\Theta}_{kL}).
\end{align}

To detect the symbol $x_k$, an arbitrary combining vector $\mathbf{d}_k\in \mathbb{C}^M$, built from $\{\hat{\mathbf{h}}_k\}_{k\in \mathbb{K}}$, is applied as $\hat{x}_k = \mathbf{d}_k^H \mathbf{y}$. This can be expanded as
\begin{equation}    
 \label{massiveMIMO:MFsoftestimateUL} 
    \hat{x}_k = \sqrt{p_u} \sum_{k'\in \mathbb{K}} \sqrt{\eta_{k'}} \mathbf{d}_k^H (\hat{\mathbf{h}}_{k'} + \tilde{\mathbf{h}}_{k'}) x_{k'} + \mathbf{d}_k^H \mathbf{z}.
\end{equation} 

\begin{proposition}
The achievable SE of user $k$ is
\begin{equation}
    R_k=\left(1-\frac{\tau_p}{\tau_u}\right)\mathbb{E} \Bigl[\log_2(1+\gamma_k) \Bigr],
\end{equation}
where the instantaneous effective signal-to-interference-plus-noise ratio (SINR) is expressed as
\begin{equation} \label{GS_SINR_UL_AP}
    \gamma_k = \frac{ \eta_k | \mathbf{d}_k^H \hat{\mathbf{h}}_k |^2  }{  \mathbf{d}_k^H \left( \sum\limits_{k'\in \mathbb{K}\setminus \{k\}} \eta_{k'}  \hat{\mathbf{h}}_{k'}\hat{\mathbf{h}}_{k'}^H  + \sum\limits_{k'\in \mathbb{K}} \eta_{k'} \boldsymbol{\Theta }_{k'} +\frac{\sigma_z^2}{p_u}\mathbf{I}_M\right) \mathbf{d}_k  }
\end{equation}
\end{proposition}
\begin{IEEEproof}
The detailed proof of this proposition is omitted here for brevity, with interested readers being directed to \cite{SIG-093} for a comprehensive derivation.
\end{IEEEproof}

\subsection{Centralized Combining Techniques}
In the following, we present various combining methods that are applied to centrally processed CF-mMIMO systems:
\begin{enumerate}
   \item \textbf{Maximal-Ratio (MR)}:  This low-complexity method sets the combining vector for user $k$ to its estimated channel: $\mathbf{d}_{k}^{mr}=\hat{\mathbf{h}}_{k}$, maximizing the desired signal while neglecting inter-user interference.
    \item \textbf{Zero-Forcing (ZF)}: Aiming to nullify inter-user interference, the ZF detector is the pseudo-inverse of the estimated channel matrix, that is, $\mathbf{A}=(\hat{\mathbf{H}}^H\hat{\mathbf{H}})^{-1}\hat{\mathbf{H}}^H$, where $\hat{\mathbf{H}} = [\hat{\mathbf{h}}_{1},\hat{\mathbf{h}}_{2}\ldots,\hat{\mathbf{h}}_{K}]$. Denoting $\mathbf{a}_k\in \mathbb{C}^{1\times M}$ as the $k^{th}$ row of $\mathbf{A}$, and the ZF combining vector for user $k$ is $\mathbf{d}_k^{zf}=\mathbf{a}_k^H$. ZF may amplify noise, especially in scenarios with low signal-to-noise ratios (SNRs).
    \item \textbf{Regularized Zero-Forcing (RZF)}: To mitigate noise amplification in low-SNR conditions, RZF introduces a regularization term $\alpha$ for improved stability. The RZF combining matrix is given by:  \begin{equation}
    \mathbf{A}_{r} = (\hat{\mathbf{H}}^H\hat{\mathbf{H}} + \alpha \mathbf{I}_K)^{-1} \hat{\mathbf{H}}^H, \end{equation} where $\alpha > 0$ is a regularization parameter that balances interference suppression and noise amplification. The RZF combining vector for user $k$ is: \begin{equation}
    \mathbf{d}_k^{rzf} = \mathbf{a}_k^H, \end{equation} where $\mathbf{a}_{k}$, in this case, is the $k$th row of $\mathbf{A}_{r}$.
    \item \textbf{MMSE}:  MMSE combining strikes a balance between amplifying the desired signal and suppressing inter-user interference plus thermal noise. By minimizing the mean squared error (MSE) between \( x_k \) and its  estimate \( \hat{x}_k = \mathbf{d}_k^H \mathbf{y} \), the MMSE combining vector is derived using the Wiener-Hopf solution: 
\[
\mathbf{d}_{k}^{mmse} = \mathbf{C}_{\mathbf{y}}^{-1} \mathbf{c}_{\mathbf{y}x_k},
\]
where $\mathbf{C}_{\mathbf{y}}=\mathbb{E}[\mathbf{y} \mathbf{y}^H \mid \{\hat{\mathbf{h}}_{k}\}] $ represents the autocorrelation matrix of \( \mathbf{y} \) and $\mathbf{c}_{\mathbf{y}x_k} = \mathbb{E}[\mathbf{y} x_k^* \mid \{\hat{\mathbf{h}}_{k}\}]$ denotes the cross-correlation vector between \( \mathbf{y} \) and \( x_k \).
These correlations are given by:
\[
   \mathbf{C}_{\mathbf{y}} = p_u \sum\nolimits_{k'\in \mathbb{K}} \eta_{k'} \left( \hat{\mathbf{h}}_{k'} \hat{\mathbf{h}}_{k'}^H + \boldsymbol{\Theta}_{k'} \right) + \sigma_z^2 \mathbf{I}_{M}
   \]
and  
\[
   \mathbf{c}_{\mathbf{y}x_k} = \sqrt{p_u \eta_k} \hat{\mathbf{h}}_{k}.
   \]
Then, we obtain \begin{equation} \label{eq:MMSE-combining}
    \mathbf{d}_{k}^{mmse} {=}   \left( p_u \sum\limits_{k'\in\mathbb{K}} \eta_{k'}\left( \hat{\mathbf{h}}_{k'} \hat{\mathbf{h}}_{k'}^{H} {+} \boldsymbol{\Theta }_{k'} \right) {+} \sigma_z^2  \mathbf{I}_{M} \right)^{-1}     \hat{\mathbf{h}}_{k}. \end{equation}
\end{enumerate}

\section{Distributed Uplink Combining}

Distributed combining offers an alternative to centralized processing, balancing performance and scalability. Each AP performs local signal processing using its channel estimates, where AP $l$ forms a soft estimate \( \hat{x}_{kl}=\mathbf{v}_{kl}^H\mathbf{y}_l \) for each \( k \in \mathbb{K} \), and transmits \( \{\hat{x}_{kl}\}_{k\in\mathbb{K}}\) to the CPU for final combining. Since channel estimation is done locally, pilot signals are not sent to the CPU. Without knowledge of channel estimates, the CPU leverages large-scale fading coefficients to  detect the signals. 

\begin{proposition}
The achievable SE for user $k$, using distributed combining, is expressed as $
    (1-\frac{\tau_p}{\tau_u}) \mathbb{E} [\log_2(1+\gamma_k^{d}) ]$,
where the instantaneous effective SINR equals to
\begin{align} \label{Gs_sinrULdistributed}
    &\gamma_k^{d} =\\ \nonumber
    & \frac{   \eta_k  \left |  \sum\nolimits_{l\in \mathbb{L}} \mathbb{E}[\mathbf{v}_{kl}^H \hat{\mathbf{h}}_{kl}] \right|^2   }{ \left\{   \begin{aligned}        
      & \sum\limits_{k'\in \mathbb{K}} \eta_{k'} \left( \sum\nolimits_{l\in \mathbb{L}}\mathbb{E}\left[\left|    \mathbf{v}_{kl}^H \hat{\mathbf{h}}_{k'l} \right|^2 \right] +\sum\nolimits_{l\in \mathbb{L}} \mathbf{v}_{kl}^H \boldsymbol{\Theta}_{k'l}  \mathbf{v}_{kl} \right) \\
      &  - \eta_{k} \left(   \sum\nolimits_{l \in \mathbb{L}} \eta_k  \left|\mathbb{E}[\mathbf{v}_{kl}^H \hat{\mathbf{h}}_{kl} ] \right|^2  \right) + \frac{\sigma_z^2}{p_u} \left ( \sum\nolimits_{l\in \mathbb{L}} \| \mathbf{v}_{kl} \|^2 \right)
    \end{aligned} \right\}    
    }.
\end{align}
\end{proposition}
\begin{IEEEproof}
The proof is omitted here for brevity, with interested readers being directed to \cite{SIG-093} for a detailed derivation.
\end{IEEEproof}

In the following, we introduce local combining methods:
\begin{enumerate}
   \item \textbf{Local MR/LSFD}: Using the MR combining, the local vector for user $k$ at AP $l$ is $\mathbf{v}_{kl}^{mr}= \hat{\mathbf{h}}_{kl}$.
    \item \textbf{Local ZF}: Each AP \( l \) estimates the channels to all \( K \) users, resulting in the local channel estimate matrix \( \hat{\mathbf{H}}_l \in \mathbb{C}^{N_a \times K} \). The local ZF combining vector for user \( k \), denoted by \( \mathbf{v}_{kl}^{\text{zf}}\), is the \( k \)-th column of \( \mathbf{V}_l^{\text{zf}} \). This matrix is the pseudo-inverse of \( \hat{\mathbf{H}}_l \), defined as \( \mathbf{V}_l^{\text{zf}} = \hat{\mathbf{H}}_l (\hat{\mathbf{H}}_l^H \hat{\mathbf{H}}_l)^{-1} \). By construction, this satisfies \( (\mathbf{V}_l^{\text{zf}})^H \hat{\mathbf{H}}_l = \mathbf{I}_K \), ensuring that inter-user interference at each AP is effectively nullified.
    \item \textbf{Local RZF}: Similarly, the local RZF detector at AP $l$ becomes $\mathbf{V}_l^{rzf}=\hat{\mathbf{H}}_l(\hat{\mathbf{H}}_l^H\hat{\mathbf{H}}_l+ \sigma_z^2\mathbf{I}_K)^{-1}$. The local RZF combining vector \( \mathbf{v}_{kl}^{\text{zf}}\) is the \( k \)-th column of \( \mathbf{V}_l^{\text{rzf}} \).
    \item \textbf{Local MMSE}:  To minimize the MSE between \( x_k \) and its local estimate \( \hat{x}_{kl} = \mathbf{v}_{kl}^H \mathbf{y}_l \) at AP $l$, the local MMSE combining vector $\mathbf{v}_{kl}^{mmse}$ is derived using the Wiener-Hopf solution: 
\[
\mathbf{v}_{kl}^{mmse} = \mathbf{C}_{\mathbf{y}_l}^{-1} \mathbf{c}_{\mathbf{y}_lx_k},
\]
where $\mathbf{C}_{\mathbf{y}_l}=\mathbb{E}[\mathbf{y}_l \mathbf{y}_l^H \mid \{\hat{\mathbf{h}}_{kl}\}] $ represents the autocorrelation matrix of \( \mathbf{y}_l \) and $\mathbf{c}_{\mathbf{y}_lx_k} = \mathbb{E}[\mathbf{y}_l x_k^* \mid \{\hat{\mathbf{h}}_{kl}\}]$ denotes the cross-correlation vector between \( \mathbf{y}_l \) and \( x_k \).
These correlations are given by:
\[
   \mathbf{C}_{\mathbf{y}_l} = p_u \sum_{k'\in \mathbb{K}} \eta_{k'} \left( \hat{\mathbf{h}}_{k'l} \hat{\mathbf{h}}_{k'l}^H + \boldsymbol{\Theta}_{k'l} \right) + \sigma_z^2 \mathbf{I}_{N_a}
   \]
and  
\[
   \mathbf{c}_{\mathbf{y}_lx_k} = \sqrt{p_u \eta_k} \hat{\mathbf{h}}_{kl}.
   \]
Then, we obtain
\begin{align} \label{eq:MMSE-combining}
&\mathbf{v}_{kl}^{mmse} = \\ \nonumber
&\left( p_u \sum\limits_{k'=1}^{K} \eta_{k'}\left( \hat{\mathbf{h}}_{k'l} \hat{\mathbf{h}}_{k'l}^{H} + \boldsymbol{\Theta }_{k'l} \right) + \sigma_z^2  \mathbf{I}_{N_a} \right)^{-1}     \hat{\mathbf{h}}_{kl}.
\end{align}
\end{enumerate}

\begin{algorithm}
\SetAlgoLined \label{Algorithm_UL}
\DontPrintSemicolon
\KwIn{$\{  \hat{\mathbf{h}}_{kl}, \boldsymbol{\Theta}_{kl}\}_{k\in\mathbb{K}, l\in\mathbb{L}}$,  $p_u$, $\sigma_z^2$, $\epsilon$; }
\KwOut{Optimal power coefficients $\{\eta_k^*\}_{k\in\mathbb{K}}$ }
Initialize: $t \leftarrow 0$, $\gamma_\text{min}^{(0)} \leftarrow 0$, $\gamma_\text{max}^{(0)} \leftarrow \max\left\{ p_u \sum_{l\in \mathbb{L}} \| \hat{\mathbf{h}}_{kl} \|^2/ \sigma_z^2\right\}$\;
\While{$\gamma_\text{max}^{(t)} - \gamma_\text{min}^{(t)} > \epsilon$}{
    $\gamma_t \leftarrow \frac{1}{2}(\gamma_\text{min}^{(t)} + \gamma_\text{max}^{(t)})$\;   
    \textbf{Convex Feasibility Check:}\; Find $\{\eta_k\}_{k\in\mathbb{K}}$\\ s.t. $\gamma_k \geq \gamma_t, 0 \leq \eta_k \leq 1$ for all $k \in \mathbb{K}$\;
    \If{feasible}{
        $\gamma_\text{min}^{(t+1)} \leftarrow \gamma_t$, $\gamma_\text{max}^{(t+1)} \leftarrow \gamma_\text{max}^{(t)}$, $\eta_k^* \leftarrow \eta_k$\;
    }
    \Else{
        $\gamma_\text{min}^{(t+1)} \leftarrow \gamma_\text{min}^{(t)}$, $\gamma_\text{max}^{(t+1)} \leftarrow \gamma_t$\;
    }
    $t \leftarrow t + 1$\;
}
\Return $\{\eta_k^*\}_{k\in\mathbb{K}}$
\caption{Universal Max-Min Power Control}
\end{algorithm}

\section{Unified Max-Min Power Optimization}
Max-min power control has been shown to significantly improve fairness in CF systems \cite{Ref_ngo2017cellfree, Ref_bashar2019uplink}. However, the algorithms were individually designed to tailor specific combining schemes in earlier works. By leveraging the unified analytical framework, we propose a universal max-min power control algorithm, which operates effectively with any combining schemes. 
The optimization formula can be represented as
\begin{equation}  
\begin{aligned} \label{eqnIRS:optimizationMRTvector}
\max_{\{\eta_k\}_{k\in\mathbb{K}}} \min_{k}\quad &  \gamma_{k}  \\
\textrm{s.t.} \quad & 0\leqslant \eta_{k} \leqslant 1,\quad \forall k\in \mathbb{K}
\end{aligned}.
\end{equation}
Introducing a slack variable $\gamma_t$, which represents the minimum SINR across users, the problem is reformulated as 
\begin{equation}  
\begin{aligned} \label{EQN_maxminOptim}
\max_{\{\eta_k\}_{k\in\mathbb{K}},\;\gamma_t } \quad &  \gamma_t \\
\textrm{s.t.} \quad & \gamma_{k} \geqslant \gamma_t, \quad \forall k\\
\quad & 0\leqslant \eta_{k} \leqslant 1,\quad \forall k\in \mathbb{K}.
\end{aligned}
\end{equation}
Since \( \gamma_{k} \) is quasiconcave with respect to \( \{\eta_k\}_{k \in \mathbb{K}} \), the constraint set \( \gamma_{k} \geq \gamma_t \) is convex. Therefore, a bisection method is applied to solve this problem, as depicted in Algorithm~\ref{Algorithm_UL}. This algorithm covers also distributed combining methods by replacing $\gamma_k$ with $\gamma_k^d$.

\section{Computational Complexity}

The computational complexity is evaluated in terms of the required number of complex multiplications. For MR combining, the detection $\hat{x}_k =  \hat{\mathbf{h}}_k^H \mathbf{y} $ requires \( M \) multiplications. For all \( K \) users, the total complexity is $K M $ complex multiplications per channel use.

The computation of the ZF combining matrix $\mathbf{A} = (\hat{\mathbf{H}}^H \hat{\mathbf{H}})^{-1} \hat{\mathbf{H}}^H$ involves three main steps:  
\begin{itemize}
    \item Compute \( \hat{\mathbf{H}}^H \hat{\mathbf{H}} \), which requires \( K^2 M \) complex multiplications.
    \item Compute the inverse \( (\hat{\mathbf{H}}^H \hat{\mathbf{H}})^{-1} \), a \( K \times K \) complex matrix, requiring approximately \( K^3 \) complex multiplications using standard methods like Lower-Upper decomposition.
    \item Multiply a \( K \times K \) matrix by a \( K \times M \) matrix for  \( \mathbf{A} = (\hat{\mathbf{H}}^H \hat{\mathbf{H}})^{-1} \hat{\mathbf{H}}^H \), requiring \( K^2 M \) complex multiplications.
\end{itemize}
Thus, it takes $2K^2 M+K^3$ multiplications per coherence block. 
Once \( \mathbf{A} \) is computed, it is used to detect the received signals for each channel use within the coherence block. This matrix-vector multiplication requires \( K M \) complex multiplications. Thus, the per-channel-user computational complexity for ZF is:
\begin{equation} \label{Gs_ZFcomplexity}
\frac{2 K^2 M + K^3}{\tau_u} + KM.
\end{equation}
For RZF, the difference from ZF is the addition of $\sigma_z^2\mathbf{I}_K$, requiring no multiplications. Thus, the complexity of RZF is the same as that of ZF. 

Building an MMSE combining vector involves computing the \( M \times M \) correlation matrix, requiring \(  K M^2 \) multiplications, inverting it at a cost of \( M^3 \), and computing combining vectors for \( K \) users, adding \( K M^2 \), totaling \( M^3 + 2 K M^2 \) multiplications per coherence block. In detection, computing inner products for $K$ users costs \( K M \) multiplications. Thus, the  complexity for the MMSE combining equals
\begin{equation}
\frac{M^3 + 2 K M^2}{\tau_u} +KM
\end{equation}
multiplications. The major difference with ZF/RZF is the matrix inverse of \( K \) and \( M \).

For distributed processing, the overall computational complexity across all $L$ APs is analyzed:
\begin{itemize}
    \item \textit{Local MR:} The complexity per AP is $K N_a$. With $L$ APs, the total complexity is $K N_a L=KM$ multiplications.    
    \item \textit{Local ZF:} Each AP first computes the Gram matrix $\hat{\mathbf{H}}_l^H \hat{\mathbf{H}}_l$ at a cost of $K^2 N_a$ multiplications, then inverts a $K \times K$ matrix at a cost of $K^3$, and finally computes the pseudo-inverse multiplication with $\hat{\mathbf{H}}_l$ at $K^2 N_a$. The detection step adds $K N_a$ multiplications per AP. With $L$ APs, the total complexity is:
    \begin{equation}
        \frac{L(2 K^2N_a  + K^3)}{\tau_u} + K LN_a=\frac{2 K^2 M + K^3L}{\tau_u} + K M.
    \end{equation}
    \end{itemize}
    \begin{remark}
\textit{
    Compared to \eqref{Gs_ZFcomplexity}, local ZF is \textbf{more complex}, as each AP performs a $K\times K$ matrix inversion, whereas in centralized ZF, the CPU only needs to perform this inversion once.}
    \end{remark}
    \begin{itemize}
       \item \textit{Local RZF:} The complexity is the same as local ZF since the additional regularization term does not contribute to extra multiplications. 
    \item \textit{Local MMSE:} Computing the autocorrelation matrix requires $K N_a^2$ multiplications per AP. Inverting the $N_a \times N_a$ matrix takes $N_a^3$, and computing the combining vectors for $K$ users adds $K N_a^2$. The detection step requires $K N_a$ multiplications. Summing up for all $L$ APs, the total complexity is:
    \begin{equation}
    \frac{MN_a^2 + 2 K MN_a }{\tau_u} +KM.
    \end{equation}
\end{itemize}

\begin{table*}[h]
    \centering
    \begin{tabular}{|c|c|c|c|}
        \hline \hline
        \textbf{Method Type} & \textbf{Combining Method} & \textbf{Computational Complexity} & \textbf{Fronthaul Overhead (Complex Scalars)} \\
        \hline
        \multirow{4}{*}{Centralized Methods} & \textbf{MR} & $ KM $ & \multirow{4}{*}{$ L N_a (\tau_p + \tau_u) $} \\
        \cline{2-3}
        & \textbf{ZF} & $ \frac{2K^2 M + K^3}{\tau_u} + KM $ &  \\
        \cline{2-3}
        & \textbf{RZF} & $ \frac{2K^2 M + K^3}{\tau_u} + KM $ &  \\
        \cline{2-3}
        & \textbf{MMSE} & $ \frac{M^3 + 2 K M^2}{\tau_u} + KM $ &  \\
        \hline
        \multirow{4}{*}{Distributed Methods} & \textbf{LSFD} & $ KM $ & \multirow{4}{*}{$  LK \tau_u $} \\
        \cline{2-3}
        & \textbf{Local ZF} & $ \frac{2 K^2 M + K^3L}{\tau_u} + K M $ &  \\
        \cline{2-3}
        & \textbf{Local RZF} & $ \frac{2 K^2 M + K^3L}{\tau_u} + K M $ &  \\
        \cline{2-3}
        & \textbf{Local MMSE} & $ \frac{MN_a^2 + 2 K MN_a }{\tau_u} + KM $ &  \\
        \hline \hline
    \end{tabular}
    \caption{Computational Complexity and Fronthaul Overhead for Centralized and Distributed Combining Methods}
    \label{tab:complexity_fronthaul}
\end{table*}

\section{Fronthaul Overhead}  

Fronthaul overhead refers to the number of complex scalars exchanged between the APs and the CPU. In centralized processing, each AP forwards the received pilot and data signals to the CPU, requiring $N_a$ complex scalars per channel use.  Consequently, the total fronthaul overhead for each coherent block is given by  
\(L \cdot N_a \cdot (\tau_p + \tau_u)\). 
In distributed processing, each AP locally computes the combined signal without sending pilot signals to the CPU. However, it forwards the detected symbols for each user to the CPU, resulting in an overall overhead of  
\(K \cdot L \cdot \tau_u\).

From a scalability perspective, we consider a practical scenario where the number of users exceeds the number of antennas per AP, i.e., \( K > N_a \), and the pilot sequence length is relatively very small compared to the uplink data transmission duration \(  \tau_p \ll \tau_u \).  Under these conditions, the fronthaul overhead for centralized processing can be approximated as $L \cdot N_a \cdot \tau_u $, while for distributed processing, it is  $K \cdot L \cdot \tau_u$. It follows that  
\[
K \cdot L \cdot \tau_u > N_a \cdot L  \cdot \tau_u.
\]  
This implies:
\begin{remark}
    \textit{Centralized combining is more efficient in terms of fronthaul overhead when \( K \) grows large, as its overhead depends on \( N_a \), which remains relatively small and fixed per AP. In contrast, distributed processing incurs \textbf{higher overhead} as it scales with the number of users, making it less efficient in large-scale deployments.}
\end{remark}

\section{Numerical Evaluation}
\begin{figure*}[!tbph]
\centering
\subfloat[]{
    \includegraphics[width=0.475\textwidth]{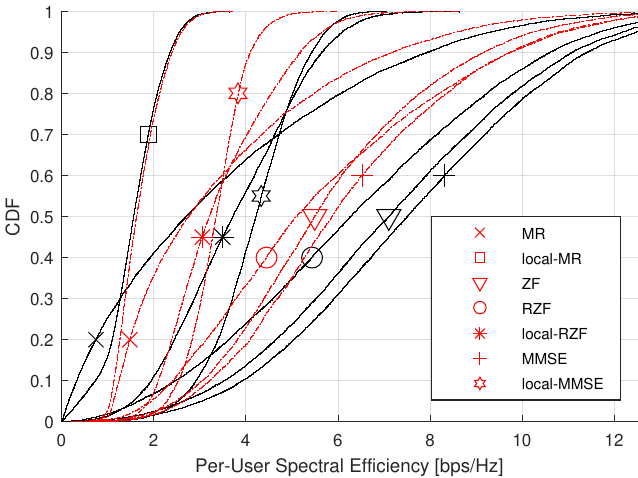}
    \label{fig:result1} 
}
\hspace{1mm}
\subfloat[]{
    \includegraphics[width=0.475\textwidth]{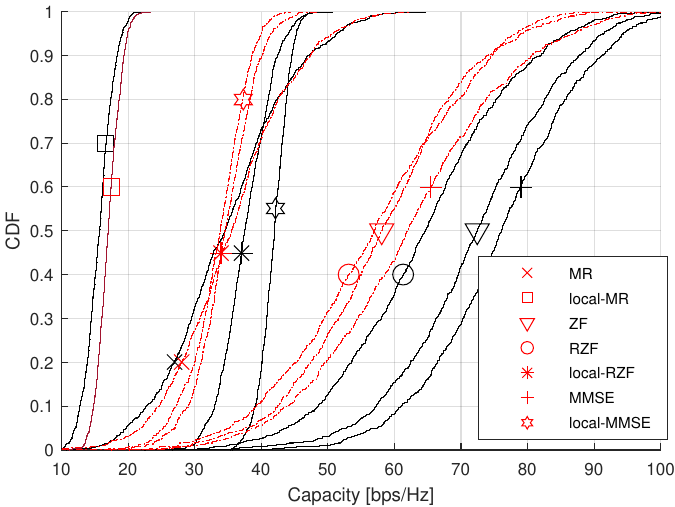}
    \label{fig:result2} 
}
\caption{Performance comparison: (a) the CDF curves in terms of per-user SE; and (b) the CDF curves of sum throughput (capacity). Solid lines indicate full power transmission, while dashed-dot lines (with matching markers) represent max-min power control.}
\label{fig:result}
\end{figure*}
In the simulations, the system is configured with the following parameters:
\begin{itemize}
    \item Total number of antennas: \( M = 256 \)
    \item Number of distributed APs: \( L = 64 \)
    \item Number of antennas per AP: \( N_a = 4 \)
    \item Number of active users: \( K = 10 \)
    \item Coverage radius: 1 km
    \item UE transmit power: \( p_u = 200 \) mW
    \item Power constraint per antenna: 50 mW, allowing each 4-antenna AP to transmit up to 200 mW
    \item Noise power spectral density: -174 dBm/Hz
    \item Noise figure: 9 dB
    \item Bandwidth: 5 MHz
    \item Antenna array configuration: half-wavelength-spaced uniform linear arrays
    \item Channel realizations: COST-Hata model (the same parameters as \cite{Ref_jiang2021cellfree}).
    \item Spatial correlation: modeled using the Gaussian local scattering model \cite{Ref_bjornson2020making} with an angular standard deviation of \( 10^\circ \)
    \item Coherence block length: \( \tau_c = 200 \) channel uses
    \item Pilot sequence length: \( \tau_p = 5 \), set to half the number of active users, leading to pilot contamination as every two users share a pilot sequence
\end{itemize}

In \figurename \ref{fig:result1}, the cumulative distribution function (CDF) plot illustrates the per-user SE results of both centralized and local combining schemes. Centralized schemes (ZF, RZF, MMSE) leverage global CSI at the CPU for coherent detection, enabling interference suppression and optimal combining. Among these, MMSE achieves the highest SE, followed closely by ZF and RZF, with ZF’s performance being notably near that of MMSE. Despite lower complexity, MR, which lacks interference suppression, performs the worst among centralized methods. In contrast, local variants (local-MR, local-RZF, local-MMSE) rely on non-coherent detection with only local CSI at each AP, leading to significantly lower SE. Local-MMSE outperforms other local schemes but remains inferior to all centralized approaches, highlighting the critical role of global CSI and coherent detection. Notably, local-ZF is absent because the underdetermined channel matrix (with AP antennas \(N_a <\) users \(K\)) does not meet the full-rank assumption and prevents matrix inversion,  rendering ZF impractical. The results further compare full power transmission (solid lines) and max-min fairness power control (dash-dot lines): power control improves fairness by boosting the SE of the weakest users. For MR, this improvement is significant due to its lack of inherent interference management, whereas for ZF and MMSE, which already mitigate interference, the gains are noticeable but less pronounced. The fairness comes with the cost of reduced peak SE, underscoring a trade-off between uniformity and peak performance.

The CDF of overall system capacity (\figurename \ref{fig:result2}) demonstrates that centralized schemes outperform local counterparts, with ZF performing notably close to MMSE. While max-min power control improves fairness by enhancing the worst-case user rates, it reduces overall system capacity by constraining peak performance, making it less favorable for capacity-centric objectives. The results underscore the trade-off between fairness and capacity: centralized processing maximizes system-wide gains, whereas power control strategies must balance priorities based on deployment goals.

\begin{figure}[!tbph]
    \centering
    \includegraphics[width=0.425\textwidth]{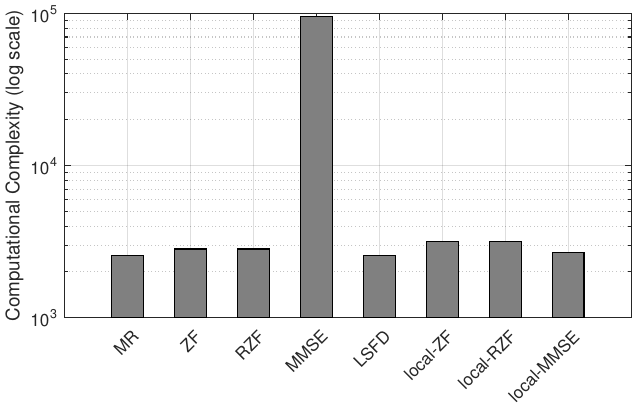}
    \caption{Comparison of computational complexity.}
    \label{fig:complexity}
\end{figure}

Regarding fronthaul overhead in our case with $K=10$ and $N_a=4$, distributed processing incurs \textbf{2.5 times as much overhead} as centralized processing. This contradicts many previous claims that local processing is more efficient than centralized processing in CF-mMIMO.
The computational complexity comparison is shown in Fig.~\ref{fig:complexity}. Centralized MMSE has the highest complexity due to the $\mathcal{O}(M^3)$ term, while local MMSE reduces complexity by leveraging distributed processing. Both MR and LSFD achieve the lowest complexity as they avoid matrix inversions. Note that local ZF/RZF exhibit higher complexity than centralized ZF/RZF due to the $K^3L$ term scaling with AP count $L$. This highlights the trade-off between centralized performance and distributed scalability.

\section{Conclusion}

To determine the most efficient technique for uplink cell-free massive MIMO, we established a unified spectral efficiency framework evaluating diverse combining schemes and conducted trade-off studies over fronthual overhead and computational complexity. Our analysis and evaluation reveal that \textbf{centralized zero-forcing (ZF) processing without power control} is the optimal choice because:
\begin{itemize}
    \item \textbf{Near-Optimal Performance:} It achieves 87\% of 5th percentile SE (fairness) of centralized MMSE in worst-case performance  and 95\% in average system capacity, while avoiding MMSE's prohibitive $\mathcal{O}(M^3)$ complexity (see Fig.~\ref{fig:complexity}). 
    \item \textbf{No power control:} It eliminates the need for max-min power control—a computationally intensive process—thereby simplifying implementation.   
    \item \textbf{Low Computational Complexity:} The complexity of centralized ZF operates at just 1.1$\times$ that of MR ($KM$), and even \textit{lower} than local ZF.   
    \item \textbf{Reduced Fronthaul Overhead:} Centralized methods require $LN_a(\tau_p + \tau_u) = 51,200$ complex scalars, compared to $LK\tau_u = 124,800$ for distributed schemes — a 59\% reduction. This contradicts prior assumptions about distributed architectures being fronthaul-efficient.
\end{itemize}

These advantages position centralized ZF as the most efficient solution for practical uplink cell-free deployments, balancing performance, complexity, and fronthaul requirements.

\bibliographystyle{IEEEtran}
\bibliography{IEEEabrv,Ref_COML}

\end{document}